# LLMs are Few-Shot Decision-Makers: Generalized Context-Aware Microgrid Frequency Control through Prompt Decision Transformer

Xu Yang, *Graduate Student Member, IEEE*, Chenhui Lin, *Senior Member, IEEE*, Haotian Liu, *Member, IEEE*, Kaihang Deng, Yunhe Li, and Wenchuan Wu, *Fellow, IEEE*

***Abstract*—The rapid evolution of energy structures has positioned microgrids as pivotal components of next-generation power systems, offering enhanced resilience and renewable energy integration. However, the inherent low inertia, complex dynamics, and poor model conditions of microgrids necessitate advanced data-driven frequency control strategies. Although reinforcement learning (RL) has demonstrated certain potential and advantages, existing RL methods often struggle with generalization across diverse microgrid configurations and lack adaptability to unseen environments, particularly when explicit system parameters are unavailable. To address these challenges, in this paper, we introduce a novel prompt decision transformer (Prompt-DT) architecture for microgrid frequency control. Unlike traditional approaches that rely on hard-to-obtain environmental characteristic parameters, the proposed method leverages few-shot expert historical trajectories as prompts to guide autonomous perception and adaptive decision-making. In addition, we propose a context-aware training and execution mechanism utilizing self-supervised contrastive learning to enhance environment recognition and prompt utilization efficiency. In addition, a physics-informed prompt design technique that filters prompts based on cumulative reward and frequency volatility is proposed, ensuring high-quality physical guidance during online execution. Finally, to ensure generalization in unseen environments with limited data, we develop a lightweight finetuning approach that achieves performance comparable to full-parameter finetuning with minimal adjustments. Extensive experiments across various microgrid configurations demonstrate that the proposed Prompt-DT significantly outperforms existing baselines in both task versatility and extrapolation capability, offering a generalized and robust solution for intelligent microgrid frequency regulation.**

## I. INTRODUCTION

DRIVEN by the transformation of the energy structure and escalating climate crises, microgrids are gradually becoming crucial units in building the next-generation power system [1], [2]. Through the coordinated operation of local generators, energy storage systems (ESSs), photovoltaics (PVs), and wind turbines (WTs), microgrids can create self-governing energy networks [3], [4]. This architecture not only boosts the local renewable energy consumption but also ensures uninterrupted power supply to critical loads under extreme events like natural disasters and main grid outages. Owing to their flexibility and resilience, microgrids stand out as pivotal technologies for clean energy transition, reliable power delivery, and system intelligence [5].

Microgrid frequency control [6], [7], serving as a critical approach for regulating system active power balance and maintaining frequency stability, is indispensable for ensuring the safe and stable operation of microgrids. Compared to transmission and distribution systems, microgrids exhibit lower inertia and narrower safety margins, along with less favorable model conditions. Therefore, more adaptive data-driven methods are required to achieve faster, more dynamic, and more precise control in modern microgrids.

Over the past few years, data-driven frequency control methods, with reinforcement learning (RL) [8] as a prime representative, have witnessed widespread research and applications [9]-[16]. For instance, researchers in [9]-[11] focused on multi-agent cooperative frequency control, utilizing multi-agent RL algorithms to achieve coordination among multiple microgrids or heterogeneous resource clusters, thereby effectively enhancing frequency regulation performance in complex situations. While researchers in [12]-[14] aimed to improve system security and robustness. By employing techniques such as meta-learning and control barrier functions, they tackle critical challenges including parameter uncertainty and defense against cyberattacks during frequency control process. Furthermore, researchers in [15], [16] targeted aggregated resources like electric vehicle aggregators and virtual power plants (VPPs), proposing hierarchical frequency control strategies to achieve a balance between economic efficiency and real-time responsiveness.

In real-world scenarios, constrained by the limited computing power and data resources at the microgrid level, local training is often unfeasible for microgrid operators. Instead, policy training is typically conducted on upper-level platforms, such as VPPs, distribution system operators, and cloud energy management systems, after which the policy networks are dispatched to individual microgrids for local execution. This "single-training and multi-site deployment"

This work was supported in part by Science and Technology Project of State Grid Corporation of China "Research on Intelligent Situation Awareness and Optimal Decision-Making Technologies for Large-Scale Urban Power Grid Operation" under Grant 52020125000X-025-ZN *(Corresponding author: Wenchuan Wu).*

Xu Yang, Chenhui Lin, Haotian Liu, Kaihang Deng, Yunhe Li, and Wenchuan Wu are with the State Key Laboratory of Power Systems, Department of Electrical Engineering, Tsinghua University, Beijing 100084, China.

paradigm imposes a demand for generalization capabilities on existing methods, which is manifested in two aspects:

1) *Task versatility.* By training in diverse environments, the model must maintain excellent control performance in each previously encountered environment. This allows a single policy network trained on the upper-level platform to be applied across different microgrids with distinct configurations, eliminating the need for site-specific retraining.

2) *Extrapolation capability.* For unseen environments, the model should maintain strong adaptability, requiring only a small number of samples to quickly learn the corresponding control strategy. This guarantees superior control performance even for newly built microgrids or those with poor data quality.

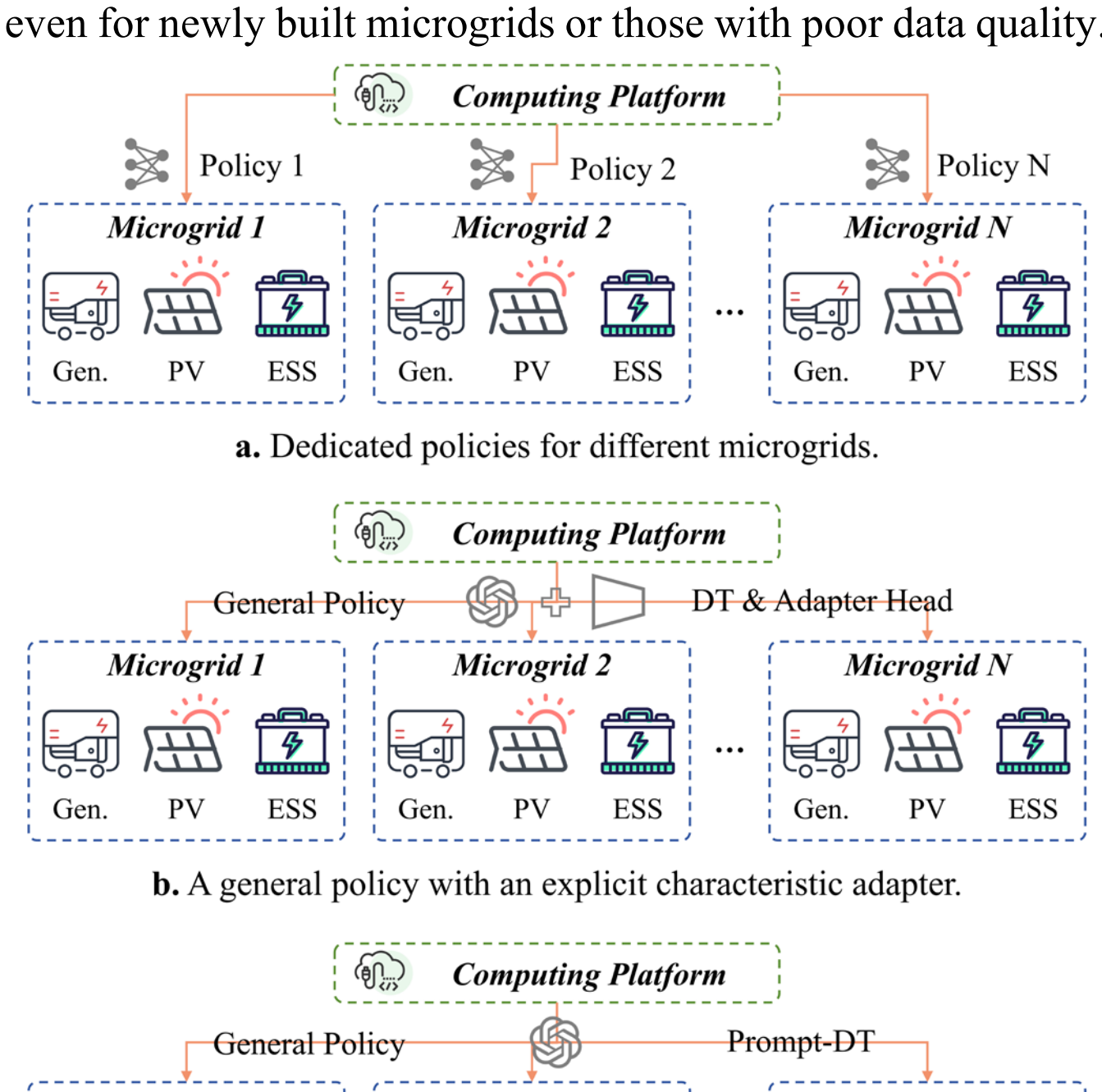

**Fig. 1.** Three different policy training paradigms for microgrid dispatch and control problems.

Despite the tremendous progress in current RL technology, generalization remains one of the main formidable barriers that existing algorithms struggle to overcome. Inspired by the powerful generalization capabilities of large language models (LLMs), some pioneering studies have begun to introduce LLMs into power system dispatch problems [17]-[25], which leverage the reasoning and comprehension abilities of LLMs to assist in strategy generation. Other studies, on the other hand, train the corresponding policy networks based on a novel LLM backbone. For example, in [26], which represents the most relevant preceding study to this work, researchers utilized a small-sized GPT-2 as the decision transformer (DT) [27] backbone for building energy management. By formulating the building energy management problem as an autoregressive sequence modeling task and training the DT across diverse building environments, the DT policy achieves excellent generalization performance, providing a novel pathway for solving optimal dispatch problems. Building upon the DT architecture, the researchers also incorporated several environmental characteristic parameters (such as the power limit and capacity of ESS) as inputs to train an independent building adapter head. By explicitly incorporating environmental characteristic parameters through the construction of the adapter head, the generalization capability of the policy network is effectively enhanced.

However, as shown in Fig. 1, in real-world applications, the critical parameters required by the adapter head are not always obtainable. In the context of microgrid frequency control, this issue is even exacerbated as numerous parameters are difficult to measure, poorly maintained, or dynamically changing. For example, from the network perspective, key parameters such as the damping coefficient and inertia time constant are hard to estimate. From the equipment perspective, the capacity of renewable energy participating in frequency control depends on user willingness, and the regulation capacity of ESS is heavily contingent upon its state of health [28]. Consequently, what microgrid operators can typically obtain is not high-quality explicit parameters, but rather near-optimal expert historical control trajectories, which necessitates that our model possesses the ability to extract implicit information from historical trajectories and acquire generalized control knowledge.

In order to address this practical demand, we extend the traditional DT into a prompt decision transformer (Prompt-DT) [29] and introduce it to the microgrid frequency control problem. Unlike the traditional DT which only models the target sequence, Prompt-DT prepends a short segment of expert trajectories as few-shot examples before the sequence to be generated, thereby providing a contextual reference for the model's control. Just as LLMs utilize user prompts to guide their content generation, Prompt-DT also incorporates this prompt segment into its decision-making process.

Nevertheless, simple prompt concatenation alone is insufficient to guarantee enhanced generalization performance. The Prompt-DT must be explicitly taught to identify and leverage the provided prompts. Therefore, we innovatively propose a context-aware Prompt-DT training and execution mechanism. Alongside the standard action head, we also construct a specific environment feature head and a learnable environment query to dynamically extract environment features, and employ self-supervised contrastive learning to adaptively distinguish between different environments. Subsequently, these environment features and processed state embeddings are jointly fed into the action head, which significantly improves the prompt utilization efficiency and environment recognition ability. Moreover, considering the inherent information density disparities among different prompts, we also propose a physics-informed prompt design technique during the execution phase to circumvent the arbitrariness associated with random sampling. By filtering prompts based on metrics of cumulative reward and frequency volatility, this ensures that the prompts provided to the Prompt-DT during execution contain richer physical information, further enhancing its online execution performance.

Finally, for the purpose of extrapolation in unseen scenarios, we propose a lightweight finetuning method, in which the LLM backbone parameters are frozen, and only the action head, environment feature head, and environment query are retained as trainable. This improves adaptability to the target environment while preserving the general capabilities of the Prompt-DT. Experimental results demonstrate that the proposed method achieves performance comparable to full-parameter finetuning with minimal computational resources, and significantly outperforms mainstream low-rank adaptation (LoRA) finetuning methods [30], making it highly suitable for microgrid scenarios with limited computing resources. Contributions of this paper can be summarized as follows:

1) **Prompt-DT network structure.** We introduce a novel Prompt-DT structure for microgrid frequency control problem. Unlike traditional methods relying on dedicated training or hard-to-obtain environmental characteristic parameters, Prompt-DT leverages few-shot expert trajectories as prompts to guide autonomous perception and adaptive decision-making. This architecture effectively bypasses the need for explicit system modeling, enabling the agent to extract implicit control knowledge from near-optimal expert historical data.

2) **Context-aware training and execution mechanism.** We develop a context-aware training mechanism utilizing self-supervised contrastive learning. By constructing a specific environment head and distinguishing between different operation scenarios, this mechanism significantly enhances Prompt-DT's environment recognition capabilities and improves the utilization efficiency of the prompt information during decision-making process. Meanwhile, the proposed physics-informed design filters prompts based on cumulative reward and frequency volatility, ensuring high-quality physical guidance in execution phase.

3) **Lightweight finetuning approach.** We propose a new lightweight finetuning approach tailored for the Prompt-DT. By freezing the parameters of the LLM backbone and only finetuning the trainable heads, it achieves efficient adaptation to unseen environments with minimal adjustments and data requirements. The proposed finetuning approach significantly enhances the practical applicability and deployment feasibility of Prompt-DT in real-world scenarios. Comprehensive comparison and ablation studies demonstrate the effectiveness and superiority of our proposed method.

## II. Preliminaries

### A. Microgrid Frequency Control Problem Formulation

In this paper, we consider microgrids equipped with local generators, ESSs, PVs, WTs, and various loads. In real-world applications, microgrids exhibit substantial heterogeneity in their configurations, such as the damping coefficient, inertia time constant, and the adjustable range of individual equipment. The linearized system dynamics of the $i$th microgrid can be expressed as follows:

$$\Delta\dot{f}_i = -\frac{D_i}{2H_i}\Delta f_i + \frac{1}{2H_i}\big(\Delta P_{m,i} + \Delta P_{ESS,i} + \Delta P_{PV,i} + \Delta P_{WT,i} - \Delta P_{d,i}\big) \quad (1)$$

$$\Delta\dot{P}_{m,i} = -\frac{1}{T_{t,i}}\Delta P_{m,i} + \frac{1}{T_{t,i}}\Delta P_{v,i} \quad (2)$$

$$\Delta\dot{P}_{v,i} = -\frac{1}{T_{g,i}}\Delta P_{v,i} - \frac{1}{R_i T_{g,i}}\Delta f_i + \frac{1}{T_{g,i}}u_{g,i} \quad (3)$$

$$\Delta\dot{P}_{ESS,i} = -\frac{1}{T_{ESS,i}}\Delta P_{ESS,i} + \frac{1}{T_{ESS,i}}u_{ESS,i} \quad (4)$$

$$\Delta\dot{P}_{PV,i} = -\frac{1}{T_{PV,i}}\Delta P_{PV,i} + \frac{1}{T_{PV,i}}u_{PV,i} \quad (5)$$

$$\Delta\dot{P}_{WT,i} = -\frac{1}{T_{WT,i}}\Delta P_{WT,i} + \frac{1}{T_{WT,i}}u_{WT,i} \quad (6)$$

$$\Delta P_{d,i} = \Delta P_{L,i} - \Delta P_{b,PV,i} - \Delta P_{b,WT,i} \quad (7)$$

where $\Delta f_i$ is the system frequency deviation; $D_i$ and $H_i$ are damping coefficient and inertia time constant, respectively; $\Delta P_{m,i}$, $\Delta P_{ESS,i}$, $\Delta P_{PV,i}$, $\Delta P_{WT,i}$, $\Delta P_{d,i}$, and $\Delta P_{v,i}$ are generator output power deviation, ESS output power deviation, PV output power deviation, WT output power deviation, net load disturbance, governor valve displacement, respectively. Specifically, the net load disturbance $\Delta P_{d,i}$ is composed of load demand disturbance $\Delta P_{L,i}$, baseline output fluctuation of PV $\Delta P_{b,PV,i}$, and baseline output fluctuation of WT $\Delta P_{b,WT,i}$. $T_{t,i}$, $T_{g,i}$, $T_{ESS,i}$, $T_{PV,i}$, and $T_{WT,i}$ are time constants of turbine, governor, ESS, PV, and WT, respectively; $R_i$ represents the droop coefficient; $u_{g,i}$, $u_{ESS,i}$, $u_{PV,i}$, and $u_{WT,i}$ are control commands for generator, ESS, PV, and WT, respectively.

When taking non-linear components in the system into account, such as the dead band of governors and the generation rate constraint of turbines, then eq. (2) and eq. (3) should be revised as:

$$\Delta\dot{P}_{m,i} = clip\left\{-\frac{1}{T_{t,i}}\Delta P_{m,i} + \frac{1}{T_{t,i}}\Delta P_{v,i}, -\sigma_{m,i}, \sigma_{m,i}\right\} \quad (8)$$

$$\Delta\dot{P}_{v,i} = \frac{u_{g,i} - \Delta P_{v,i}}{T_{g,i}} - \frac{\max\{0, \Delta f_i - db_i\} + \min\{0, \Delta f_i + db_i\}}{R_i T_{g,i}} \quad (9)$$

where $\sigma_{m,i}$ is the generation rate limit; $db_i$ is the dead band width of the governors. Here, the clip function constrains the variation rate of $\Delta P_{m,i}$ within the bounds of $[-\sigma_{m,i}, \sigma_{m,i}]$.

### B. Sequence Modeling of Microgrid Frequency Control

Similar to RL, both traditional DT and Prompt-DT are based on the Markov decision process (MDP) [31] framework for policy optimization or expert policy approximation. Considering the characteristics of microgrid frequency control problem, we design the key components of corresponding MDP for the $i$th microgrid as follows:

1) *State space:* The state variable $s_t$ indicates the state of the environment at time step $t$, encompassing the frequency status of the system and the power output status of each equipment:

$$s_t = \big(\Delta f_{i,t}, \Delta P_{m,i,t}, \Delta P_{v,i,t}, \Delta P_{ESS,i,t}, \Delta P_{PV,i,t}, \Delta P_{WT,i,t}\big) \quad (10)$$

2) *Action space:* The action variable $a_t$ represents the action generated by the policy network at time step $t$, encompassing all control commands:

$$a_t = \big(u_{g,i,t}, u_{ESS,i,t}, u_{PV,i,t}, u_{WT,i,t}\big) \quad (11)$$

3) *Reward function:* The reward $r_t$ indicates the control effectiveness at time step $t$, whose objective is to minimize system frequency fluctuations and deviations while avoiding violation conditions, thereby preventing severe consequences such as renewable tripping or system blackouts:

$$r_t = -\beta_1(\Delta f_{i,t})^2$$
$$-\beta_2(\max\{0, \Delta f_i - \Delta f_{max}\} + \min\{0, \Delta f_i + \Delta f_{max}\})^2 \quad (12)$$

where $\Delta f_{max}$ is the allowed maximum frequency deviation; $\beta_1$ and $\beta_2$ are corresponding weight coefficients for frequency deviations and violations.

It should be noted that the MDP modeling in this paper is relatively simplified. To better characterize the system dynamics or achieve multi-objective optimization, additional elements can be smoothly incorporated into the MDP definition. For instance, indicators such as area control error can be added to the state space to reflect the system's historical states and future frequency trends. Similarly, control costs can also be integrated into the reward function to penalize unnecessary control actions. We adopt a simplified MDP modeling here because the current definition is sufficient to characterize the corresponding optimization problem.

Different from RL, which relies on the Markov property of MDPs and dictates actions $a_t$ based exclusively on the current state $s_t$, DT and Prompt-DT prioritize complete historical trajectories and long-term returns. By formulating the RL optimization problem as a conditional sequence modeling task, they harness the Transformer's attention mechanism [32] to realize autoregressive prediction of optimal actions. In this architecture, the return-to-go (RTG), i.e., the future cumulative return, is introduced into the input sequence as a key conditioning variable instead of the immediate reward, guiding the policy to generate behaviors aligned with specific return expectations. The RTG at time step $t$ $RTG_t$ is calculated as follows:

$$RTG_t = \sum_{t'=t}^{T} r_{t'} \quad (13)$$

where $T$ is the length of the frequency control process, and we set it as an hour in this study. In this way, the complete sequence comprising RTGs, state variables, and action variables is uniformly modeled as a trajectory. This formulation effectively avoids the error propagation problem inherent in traditional value iteration, and significantly enhances policy stability and performance by leveraging long-term dependency capture capabilities.

In terms of the training mechanism, departing from the trial-and-error exploration mode of RL, DT and Prompt-DT collect expert control trajectories as training data and fit the expert policy through supervised training. When a decision is required at time step $t$, the sequence prior to $t$ is fed into the model to autoregressively predict the next action.

## III. Methods

### A. Prompt-DT Network Structure

To achieve efficient sequence modeling, both DT and Prompt-DT utilize the Transformer architecture to model and capture the relationships among RTGs, states, and actions. The network structure, forward propagation process, and context-aware training mechanism of the proposed Prompt-DT are illustrated in Fig. 2.

1) *Token encoders*

As we mentioned previously, environmental characteristic parameters in real-world scenarios may not be explicitly obtainable, and only a few expert historical trajectories can be acquired. This is one of the primary reasons for introducing Prompt-DT. The most significant difference from the traditional DT is that the input of Prompt-DT not only contains the target sequence $\tau^{tgt}$, but also prepends a segment of expert trajectories as a prompt $\tau^{pmt}$ to form the complete input $\tau = (\tau^{pmt}, \tau^{tgt})$:

$$\tau^{pmt} = \left(RTG_0^{pmt}, s_0^{pmt}, a_0^{pmt}, \dots, RTG_{k^{pmt}}^{pmt}, s_{k^{pmt}}^{pmt}, a_{k^{pmt}}^{pmt}\right)$$
$$\tau^{tgt} = \left(RTG_0^{tgt}, s_0^{tgt}, a_0^{tgt}, \dots, RTG_{k^{tgt}}^{tgt}, s_{k^{tgt}}^{tgt}, a_{k^{tgt}}^{tgt}\right) \quad (14)$$

where $k^{pmt}$ and $k^{tgt}$ are the length of the prompt and the target sequence, respectively, resulting in the total length of the entire input is $3(k^{pmt} + k^{tgt})$.

Subsequently, several multi-layer perceptron (MLP) encoders project these inputs into a vector space of the same dimension. The prompt and the target sequence share the identical RTG encoder $\mathrm{Enc}^{RTG}(\cdot)$, state encoder $\mathrm{Enc}^{s}(\cdot)$, action encoder $\mathrm{Enc}^{a}(\cdot)$, and positional embedding $\mathrm{PE}(\cdot)$. For the $j$th input in the prompt, its RTG token $\widehat{RTG}_j^{pmt}$, state token $\hat{s}_j^{pmt}$, and action token $\hat{a}_j^{pmt}$ are encoded as follows:

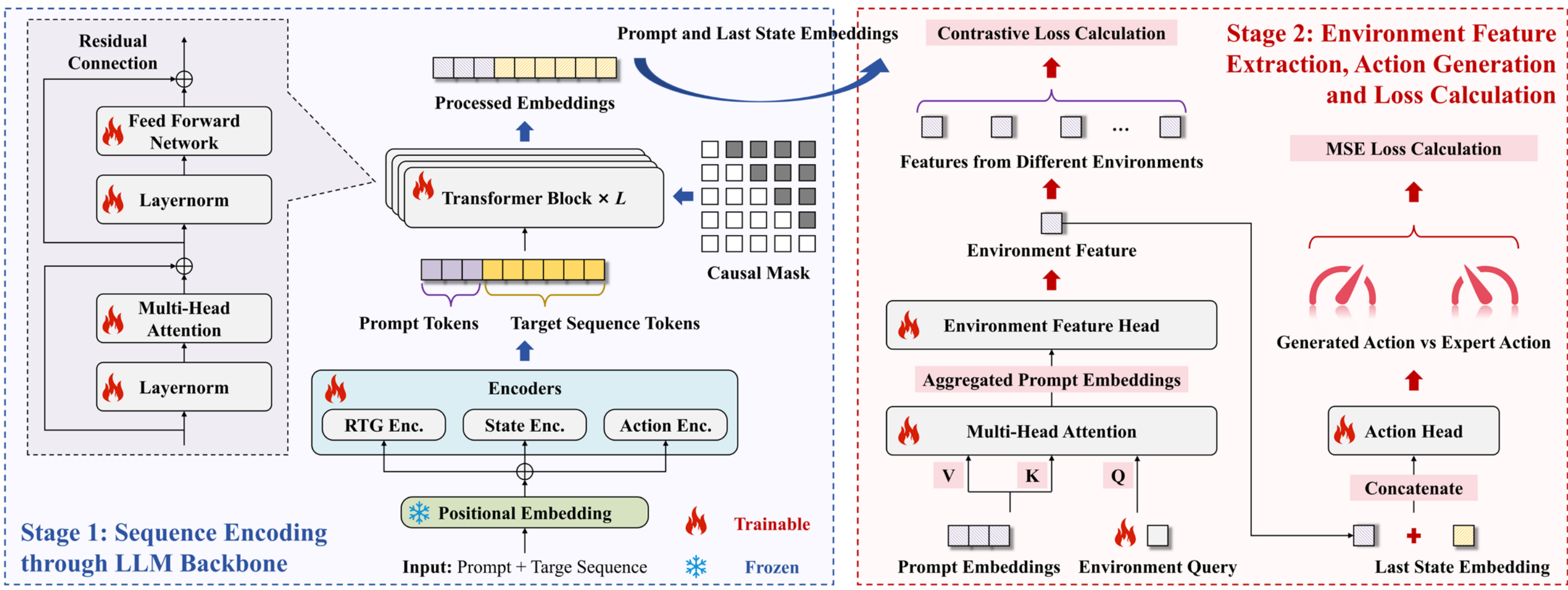

**Fig. 2.** Prompt-DT network structure and proposed context-aware training mechanism.

$$\widehat{RTG}_j^{pmt} = \mathrm{Enc}^{RTG}\left(RTG_j^{pmt}\right) + \mathrm{PE}(3j)$$
$$\hat{s}_j^{pmt} = \mathrm{Enc}^{s}\left(s_j^{pmt}\right) + \mathrm{PE}(3j+1)$$
$$\hat{a}_j^{pmt} = \mathrm{Enc}^{a}\left(a_j^{pmt}\right) + \mathrm{PE}(3j+2) \tag{15}$$

Similarly, for the $j$th input in the target sequence, its RTG token $\widehat{RTG}_j^{tgt}$, state token $\hat{s}_j^{tgt}$, and action token $\hat{a}_j^{tgt}$ are encoded as follows:

$$\widehat{RTG}_j^{tgt} = \mathrm{Enc}^{RTG}\left(RTG_j^{tgt}\right) + \mathrm{PE}(3k^{pmt} + 3j)$$
$$\hat{s}_j^{tgt} = \mathrm{Enc}^{s}\left(s_j^{tgt}\right) + \mathrm{PE}(3k^{pmt} + 3j + 1)$$
$$\hat{a}_j^{tgt} = \mathrm{Enc}^{a}\left(a_j^{tgt}\right) + \mathrm{PE}(3k^{pmt} + 3j + 2) \tag{16}$$

The complete encoded tokens are $\hat{\tau} = (\hat{\tau}^{pmt}, \hat{\tau}^{tgt})$, in which:

$$\hat{\tau}^{pmt} = \left(\widehat{RTG}_0^{pmt}, \hat{s}_0^{pmt}, \hat{a}_0^{pmt}, \dots, \widehat{RTG}_{k^{pmt}}^{pmt}, \hat{s}_{k^{pmt}}^{pmt}, \hat{a}_{k^{pmt}}^{pmt}\right)$$
$$\hat{\tau}^{tgt} = \left(\widehat{RTG}_0^{tgt}, \hat{s}_0^{tgt}, \hat{a}_0^{tgt}, \dots, \widehat{RTG}_{k^{tgt}}^{tgt}, \hat{s}_{k^{tgt}}^{tgt}, \hat{a}_{k^{tgt}}^{tgt}\right) \tag{17}$$

2) *LLM backbone*

After passing through the encoders, the tokens are then fed into the LLM backbone, which is constructed by stacking multiple Transformer blocks. Within each Transformer block, the tokens $\hat{\tau}$ undergo the following processing procedure:

$$\hat{\tau}' = \hat{\tau} + \mathrm{Attn}\left(\mathrm{LN}(\hat{\tau})\right) \tag{18}$$

Here, the residual connection is utilized to prevent gradient vanishing and network degradation; $\mathrm{LN}(\cdot)$ is the layer normalization, employed to stabilize training and accelerate convergence; $\mathrm{Attn}(\cdot)$ denotes a multi-head attention layer, whose output is computed based on the query vector $Q$, key vector $K$, and value vector $V$:

$$\mathrm{Softmax}\left(\frac{QK^T}{\sqrt{d_k}} + \mathrm{Mask}\right)V \tag{19}$$

where $d_k$ is the dimensionality of the vector; $\mathrm{Softmax}(\cdot)$ represents the softmax function; Mask is a causal mask designed to ensure that the Prompt-DT can only attend to tokens preceding the current one, thereby preventing information leakage.

Then, a feed forward network $\mathrm{FFN}(\cdot)$ and another residual connection are applied to complete the processing of this block:

$$\hat{\tau}'' = \hat{\tau}' + \mathrm{FFN}\left(\mathrm{LN}(\hat{\tau}')\right) \tag{20}$$

Following the LLM backbone, the encoded tokens are transformed into processed embeddings $\tilde{\tau} = (\tilde{\tau}^{pmt}, \tilde{\tau}^{tgt})$, with each embedding vector incorporating information from other tokens via the attention mechanism. Consequently, the target sequence embeddings $\tilde{\tau}^{tgt}$ also encapsulate the prompt information, which facilitates more accurate decision-making for the Prompt-DT.

3) *Action head*

Finally, the state embedding $\tilde{s}_{k^{tgt}}^{tgt}$ at the last position of $\tilde{\tau}^{tgt}$ is retrieved and fed into an action head, which converts it into an action $a_{k^{tgt}}^{pred}$ that conforms to the action space:

$$a_{k^{tgt}}^{pred} = \mathrm{AH}\left(\tilde{s}_{k^{tgt}}^{tgt}\right) \tag{21}$$

where the action head $\mathrm{AH}(\cdot)$ is also an MLP. In this way, the mean squared error (MSE) between the predicted action $a_{k^{tgt}}^{pred}$ and the expert historical action $a_{k^{tgt}}^{real}$ can be calculated, allowing the Prompt-DT to approximate the experts' policy by minimizing the MSE loss:

$$\mathcal{L}^{MSE} = \frac{1}{N}\sum_{i=1}^{N}\left(a_{k^{tgt},i}^{pred} - a_{k^{tgt},i}^{real}\right)^2 \tag{22}$$

where $N$ is the training batch size.

*B. Context-Aware Training Mechanism*

1) *Environment feature extraction*

The above outlines the processing pipeline of the existing Prompt-DT. It can be observed that it relies entirely on the attention mechanism to incorporate prompt information during decision-making. However, simple prompt concatenation alone is insufficient. The model must also be taught to extract environmental features, utilize them effectively, and distinguish the features from different environments, so as to explicitly account for environmental properties.

Therefore, in addition to the standard action head, we also design a learnable environment query $Q^{env}$ and an MLP environment feature head. After the prompt and target sequence are processed by the encoders and the LLM backbone, we extract the prompt embeddings $\tilde{\tau}^{pmt}$. These embeddings, along with a learnable environment query $Q^{env}$, are fed into a multi-head attention module. Here, $\tilde{\tau}^{pmt}$ serves as the keys and values, while $Q^{env}$ acts as the query, thereby aggregating the $k^{pmt}$-length prompt embeddings into a single vector of length one. After the aggregation is completed, this vector is processed by an environment feature head $\mathrm{EFH}(\cdot)$ to extract the prompt-specific environment feature $x^{env}$. Afterwards, $x^{env}$ and $\tilde{s}_{k^{tgt}}^{tgt}$ are concatenated and passed into the action head:

$$a_{k^{tgt}}^{pred} = \mathrm{AH}\left(\mathrm{Concat}\left(x^{env}, \tilde{s}_{k^{tgt}}^{tgt}\right)\right) \tag{23}$$

This enables the action head to explicitly incorporate diverse environmental characteristics into its feed forward process, instead of depending exclusively on the LLM backbone. Also, compared to the adapter head in [26], these environment features are autonomously extracted from the prompts by the model, rather than being explicitly provided as environmental characteristic parameters.

2) *Self-supervised contrastive learning*

Although such operations significantly enrich decision-making information, it simultaneously poses two critical issues, i.e., where the training signal for $x^{env}$ originates, and how to instruct the model to distinguish the $x^{env}$ representations of different environments. Considering these two challenges, we innovatively design a self-supervised contrastive learning method to assist in Prompt-DT training. Its fundamental principle is that the environment features $x^{env}$ extracted from the same environment should be more similar, while the distance between those from distinct environments should be maximized.

Based on this idea, to effectively pull similar environment features closer and push distinct ones further apart, we adopt an information noise contrastive estimation loss function. Specifically, within a training batch, we first apply L2 normalization to the extracted environment feature $x^{env}$ and compute the cosine similarity matrix among them. Then, a binary mask is constructed using the environment labels to distinguish positive pairs (same environment) from negative pairs (different environments). Next, the sum of normalized exponential similarities is calculated to derive the log probability for each sample. And finally, the self-supervised contrastive loss is obtained by averaging the negative log-likelihoods of all positive pairs, which is formulated as follows:

$$\mathcal{L}^{cont} = -\frac{1}{N}\sum_{i=1}^{N}\frac{1}{|\mathcal{P}(i)|}\sum_{j\in\mathcal{P}(i)}\log\frac{\exp\left(x_i^{env}\cdot x_j^{env}/temp\right)}{\sum_{k=1,k\neq i}^{N}\exp\left(x_i^{env}\cdot x_k^{env}/temp\right)} \tag{24}$$

where $\mathcal{P}(i)$ denotes the set of positive samples belonging to the same environment as sample $i$, and $temp$ is the temperature parameter. By minimizing this contrastive loss, the distances between positive samples are effectively reduced while negative samples are separated, which ultimately improves the model's environmental perception and recognition capabilities. The complete loss function for the Prompt-DT is as follows:

$$\mathcal{L}^{total} = \mathcal{L}^{MSE} + \lambda^{cont}\mathcal{L}^{cont} \tag{25}$$

where $\lambda^{cont}$ is the weight coefficient for the self-supervised contrastive loss. Meanwhile, we employ a dynamic scheduling strategy for $\lambda^{cont}$ during training. By initially stabilizing the MSE loss before shifting focus to the contrastive loss, we avoid the training instability that may arise from a dominant contrastive loss at the beginning.

### C. Physics-Informed Prompt and Lightweight Finetuning

1) *Physics-informed prompt design technique*

After the context-aware training, one of the most important issues during the execution phase is how to select the prompt provided to the Prompt-DT, as the information content embedded in different prompts varies significantly. Taking microgrid frequency control as an example, if the selected prompt corresponds to a highly steady control process, the model will struggle to capture key system characteristics, such as the damping coefficient, inertia time constant, equipment controllable range, and equipment response performance. Similarly, if the selected prompt segment reflects poor expert behavior, it also lacks reference value and may instead mislead the model's inference.

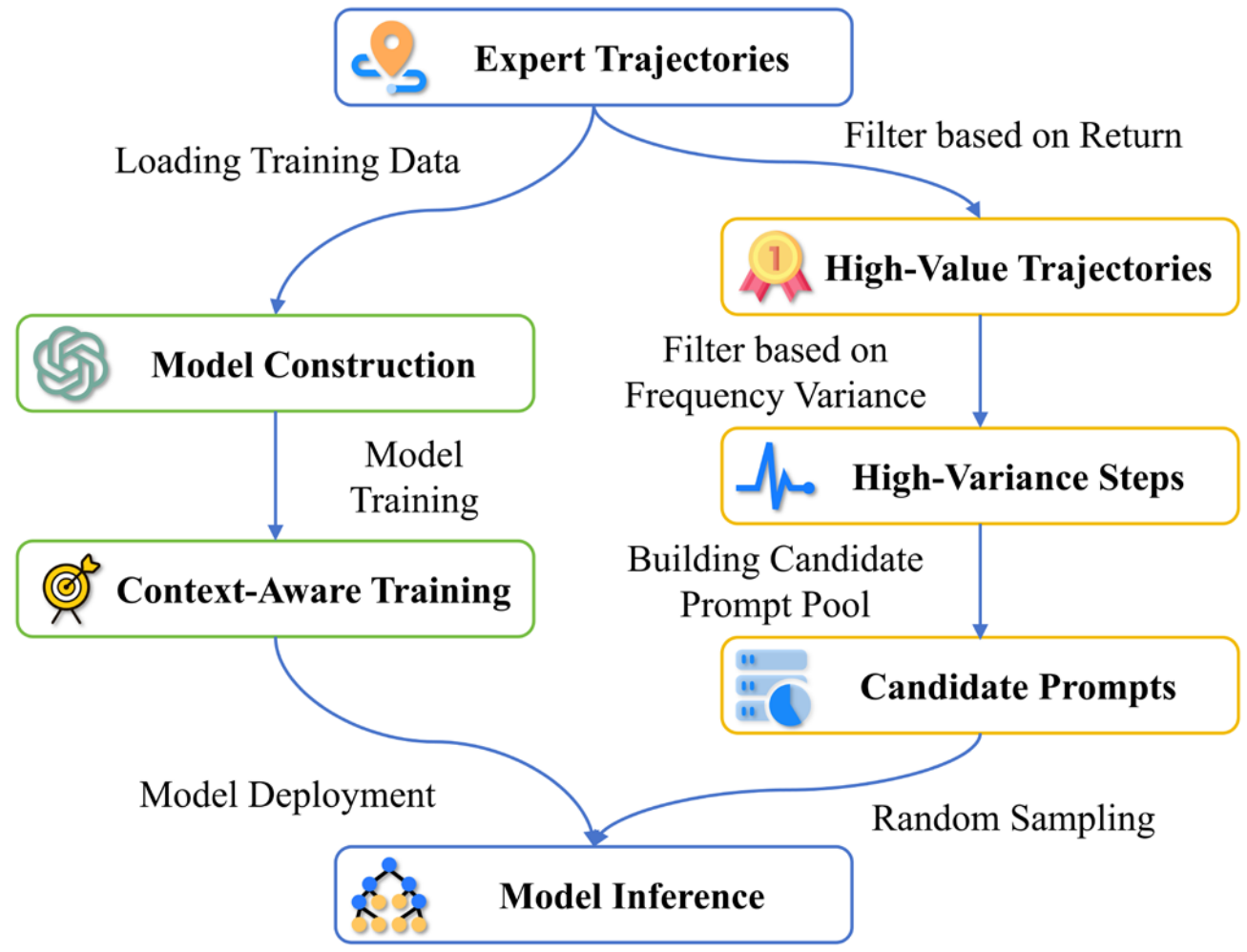

**Fig. 3.** Physics-informed prompt design process.

To overcome the drawbacks of randomly selecting prompts, we propose a physics-informed prompt design technique, which not only provides the model with high-value references but also encapsulates rich physical information. The overall process is presented in Fig. 3.

As can be seen from Fig. 3, the expert historical control trajectories are first loaded, and high-value trajectories are selected based on their cumulative rewards, as they represent high-quality expert decisions. Next, we evaluate the local frequency variance within these trajectories to identify the most dynamic time steps. This is because segments with significant frequency fluctuations typically encapsulate the richest physical information regarding system dynamics. Besides, the decision-making performance under complex steps better reflects the true proficiency of expert policy. Following this, a candidate prompt pool is constructed. Then, during the inference phase, prompts are randomly sampled from this pool to ensure diverse environmental conditioning. Compared to random selection, this physics-informed prompt design technique provides a greater number of high-quality prompts that embed richer physical information.

2) *Lightweight finetuning approach*

Through pretraining, Prompt-DT acquires the ability to approximate expert policies in various environments and can be directly applied to previously encountered environments. Thus, the policy trained on the upper-level platform can be deployed in diverse microgrids simply by prepending the expert trajectory of the corresponding environment as a prompt.

However, to apply the model in unseen environments, the pretrained Prompt-DT still requires finetuning to ensure its control effectiveness in the target environment. In this scenario, the adaptability of Prompt-DT is also demonstrated, which can achieve highly efficient control with only a small number of samples, avoiding the trial-and-error process required by RL algorithms. As shown in Fig. 4, tailored to the architectural characteristics of the Prompt-DT network, we proposed a novel finetuning approach. Different from mainstream LoRA and full-parameter finetuning, we freeze the parameters of the encoders and LLM backbone and exclusively adjust the environment query, environment feature head, and action head. This strategy achieves scenario adaptation with minimal adjustments while maximally preserving the general capabilities of Prompt-DT. As for the loss function during finetuning phase, since the finetuning is specifically tailored to the target environment, the loss function in this stage consists solely of the MSE loss $\mathcal{L}^{MSE}$ calculated in eq. (22).

The advantages of the proposed method are threefold. First, it requires a minimal number of parameter adjustments. Compared to full-parameter finetuning, the proposed lightweight approach only adjusts approximately 10% of the total parameters. Furthermore, since it does not require modifying the LLM backbone, it is even more lightweight than LoRA methods, significantly reducing the computational cost during finetuning process. Second, it only requires a small amount of data. Because the pretrained Prompt-DT inherently encapsulates general frequency control knowledge, only a limited amount of finetuning data is needed to achieve expert-level policies. This effectively improves usability in new scenarios where data may be scarce, such as newly built microgrids. Third, it exhibits high reusability. Since the LLM backbone parameters remain unaltered and unpolluted, the model can be reused in future microgrids, avoiding the need for complete retraining.

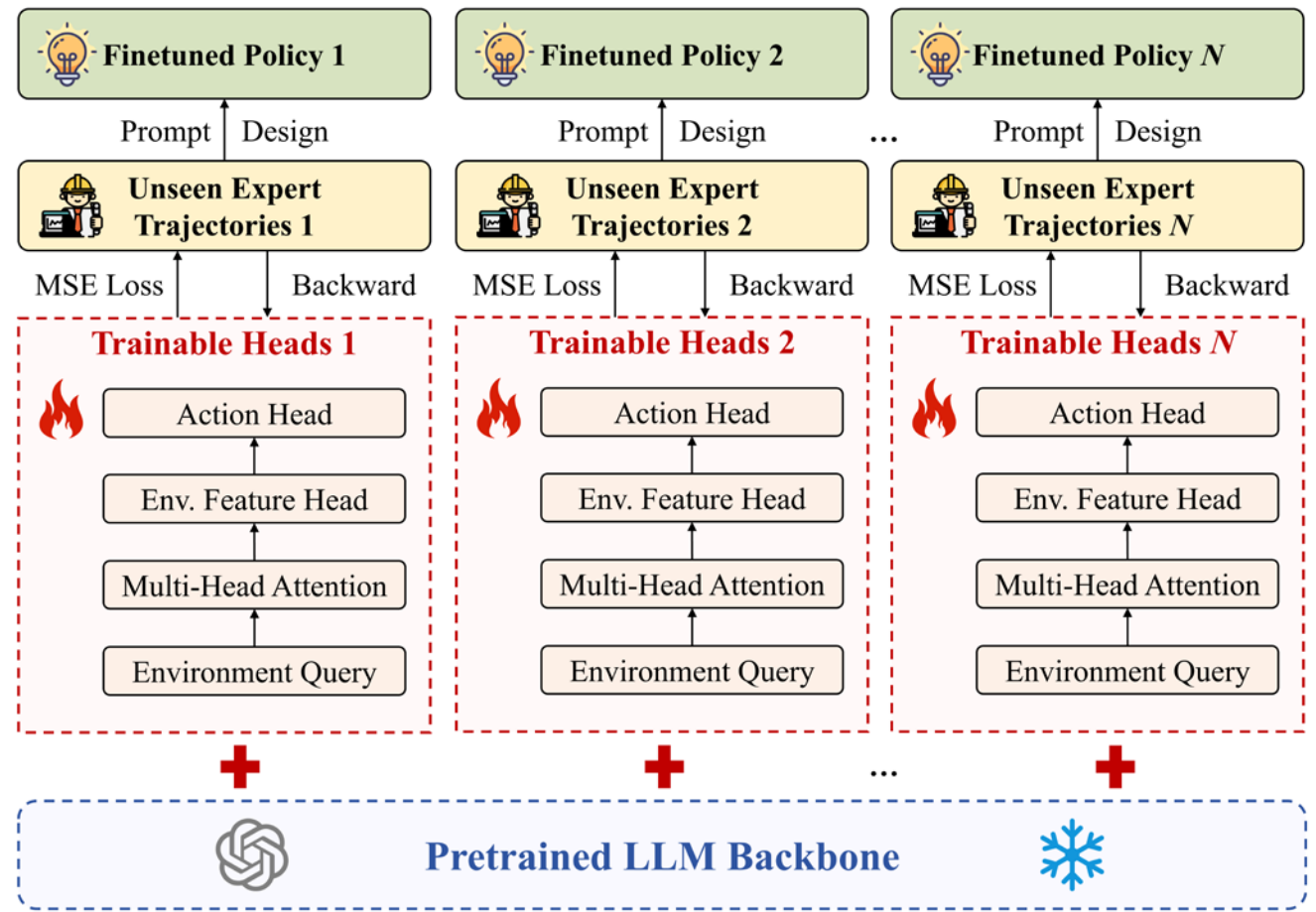

**Fig. 4.** Lightweight finetuning approach for different unseen environments.

Finally, it should be noted that this paper focuses on the microgrid frequency control problem mainly due to its dynamics and environment-specific characteristics. However, the proposed method is also applicable to other scenarios, such as microgrid energy management and microgrid participation in power markets, requiring only corresponding adjustments to the MDP formulation.

## IV. Numerical Studies

### A. Cases and Methods Setup

In this section, in order to verify the effectiveness of the proposed method, we set up 9 distinct microgrid frequency control environments. Each environment features different configurations, parameters, and equipment regulation capacities. The first 8 environments are used to train the policy networks to evaluate the task versatility of the models, i.e., whether they can adaptively recognize encountered environments and achieve optimal control performance. The remaining environment is used to finetune the pretrained Prompt-DT, validating the effectiveness of the proposed lightweight finetuning approach. For the frequency control problem, we set the length of one episode as an hour, and the control interval is 4 seconds. Nominal frequency is 50 Hz, and frequency limitations are set to [49.9, 50.1] Hz. Due to space limitations, the complete specifications and configurations of the 9 microgrids are provided in an online supplementary file [33]. It should be noted that to increase the difficulty of the problem, the second environment is configured as a challenging low-inertia microgrid, and its performance results are shaded in gray in the following subsections.

Next, to simulate the expert policy in each environment, we employ the proximal policy optimization (PPO) [34] algorithm to conduct individual training in each environment. The converged policy is then used as the expert policy. Hyperparameters for training the expert PPO are presented in Table I. After obtaining the expert policies, we deploy them in their respective environments to collect 240 hours of microgrid operation data as expert historical trajectories for subsequent model training. Hyperparameters of our proposed Prompt-DT with context-aware training are listed in Table II. As can be seen, due to the adoption of a small-sized LLM backbone, the overall trainable parameter count of the Prompt-DT is only 1,332,356, enabling training on a single GPU or a personal computer. In this paper, all experiments are conducted on a personal laptop equipped with an NVIDIA GeForce RTX 4070 Laptop GPU with 8 GB of memory.

TABLE I
HYPERPARAMETERS FOR EXPERT PPO TRAINING

| Hyperparameter | Value |
|---|---|
| Mini Batch Size | 256 |
| Hidden Layers | 2 |
| Hidden Units | 256 |
| Discount Factor | 0.9 |
| Learning Rate | 5.0e-4 → 1.0e-5 (Linear Decay) |
| Training Episodes | 1500 |

TABLE II
HYPERPARAMETERS OF THE PROPOSED METHOD

| Hyperparameter | Value |
|---|---|
| LLM Backbone | GPT-2 |
| Hidden Units | 128 |
| Transformer Layers | 6 |
| Attention Heads | 4 |
| Prompt Length | 10 |
| Target Sequence Length | 10 |
| Dropout | 0.1 |
| Heads Hidden Layers | 2 |
| Batch Size | 64 |
| Learning Rate | 1.0e-4 (Warmup + Cosine Annealing) |
| Training Episodes | 200 |
| Early Stopping Patience | 50 |
| Gradient Clip Norm | 1.0 |
| $temp$ | 0.1 |
| $\lambda^{cont}$ | 1.0e-4 (Linear Rise during Warmup) |
| Total Parameters | 1,332,356 |

In addition, to demonstrate the superiority of the proposed method, we also set up several baseline methods for comparison during the training phase: 1) Online RL (On-RL): the PPO algorithm is utilized for mixed training across the 8 training environments; 2) Offline-RL (Off-RL): the soft actor-critic (SAC) [35] algorithm is utilized to train on the 8 × 240 hours expert historical trajectories; 3) Behavior cloning (BC): the imitation learning is applied using the 8 × 240 hours expert trajectories; 4) DT: the traditional DT; 5) Prompt-DT: the proposed Prompt-DT without context-aware training mechanism, i.e., with the environment feature extraction and self-supervised contrastive learning removed; 6) Expert policy: the expert PPOs are also introduced as a baseline to provide a reference for all methods. Hyperparameters of these baseline methods are also provided in the online supplementary file [33].

### B. Performance on Encountered Environments

After training the proposed and baseline methods using their corresponding algorithms, we obtain their respective policy networks. These policies are then deployed in the first 8 microgrid environments for a 48-hour testing period to evaluate their frequency control performance in encountered environments, as presented in Table III. Specifically, Table III-A illustrates the step-wise frequency deviations, while Table III-B displays the hourly frequency violations. Policy with the best performance (excluding the expert policy) is marked in bold.

TABLE III

MICROGRID FREQUENCY CONTROL PERFORMANCE OF DIFFERENT METHODS

A. FREQUENCY DEVIATIONS ACROSS THE TEST 48 HOURS MICROGRID OPERATION (HZ)

| Env. | On-RL | | Off-RL | | BC | | DT | | Prompt-DT | | Proposed | | Expert | |
|---|---|---|---|---|---|---|---|---|---|---|---|---|---|---|
| ID | Mean | Std. | Mean | Std. | Mean | Std. | Mean | Std. | Mean | Std. | Mean | Std. | Mean | Std. |
| 1 | 6.10e-3 | 5.11e-3 | 8.40e-2 | 2.93e-2 | 6.83e-2 | 2.54e-2 | 7.81e-3 | 7.78e-3 | **3.42e-3** | 3.66e-3 | 3.47e-3 | **3.65e-3** | 2.84e-3 | 3.29e-3 |
| 2 | 2.01e-2 | 1.59e-2 | 6.52e-2 | 2.85e-2 | 5.25e-2 | 1.91e-2 | 3.65e-2 | 4.66e-2 | 3.46e-2 | 5.27e-2 | **9.22e-3** | **1.48e-2** | 3.07e-3 | 3.56e-3 |
| 3 | 1.81e-2 | 1.35e-2 | 1.16e-1 | 4.08e-2 | 9.03e-2 | 3.42e-2 | 2.90e-3 | 3.48e-3 | 3.08e-3 | **2.45e-3** | **2.68e-3** | 2.47e-3 | 2.25e-3 | 2.14e-3 |
| 4 | 1.21e-2 | 8.78e-3 | 8.91e-2 | 2.41e-2 | 6.81e-2 | 2.77e-2 | **1.60e-3** | 2.64e-3 | 1.63e-3 | **2.27e-3** | 1.65e-3 | 2.50e-3 | 1.44e-3 | 2.37e-3 |
| 5 | 1.49e-2 | 1.16e-2 | 7.92e-2 | 3.80e-2 | 6.65e-2 | 2.07e-2 | 6.88e-3 | 1.10e-2 | **4.38e-3** | **5.72e-3** | 4.93e-3 | 6.30e-3 | 3.87e-3 | 4.74e-3 |
| 6 | 1.22e-2 | 8.56e-3 | 9.00e-2 | 4.60e-2 | 6.60e-2 | 2.68e-2 | 4.51e-3 | 1.05e-2 | 4.00e-3 | 3.80e-3 | **3.14e-3** | **3.34e-3** | 2.04e-3 | 2.68e-3 |
| 7 | 1.43e-2 | 1.11e-2 | 9.81e-2 | 2.65e-2 | 7.47e-2 | 3.55e-2 | 1.36e-2 | 1.56e-2 | **2.83e-3** | **3.05e-3** | 3.01e-3 | 3.20e-3 | 1.92e-3 | 2.59e-3 |
| 8 | 7.67e-3 | 5.74e-3 | 8.07e-2 | 2.69e-2 | 6.33e-2 | 2.48e-2 | 8.07e-3 | 9.44e-3 | **3.30e-3** | 4.00e-3 | 3.74e-3 | **3.94e-3** | 2.36e-3 | 2.75e-3 |

B. FREQUENCY VIOLATIONS ACROSS THE TEST 48 HOURS MICROGRID OPERATION (HZ)

| Env. | On-RL | | Off-RL | | BC | | DT | | Prompt-DT | | Proposed | | Expert | |
|---|---|---|---|---|---|---|---|---|---|---|---|---|---|---|
| ID | Mean | Std. | Mean | Std. | Mean | Std. | Mean | Std. | Mean | Std. | Mean | Std. | Mean | Std. |
| 1 | **0.00** | **0.00** | 2.95e-1 | 8.64e-2 | 4.43e-2 | 2.63e-2 | 6.39e-4 | 3.29e-3 | **0.00** | **0.00** | **0.00** | **0.00** | 0.00 | 0.00 |
| 2 | **7.15e-4** | **2.21e-3** | 2.92e-1 | 1.53e-1 | 3.39e-2 | 3.13e-2 | 1.06e-1 | 6.91e-2 | 5.45e-2 | 3.80e-2 | 1.08e-2 | 1.63e-2 | 0.00 | 0.00 |
| 3 | **0.00** | **0.00** | 4.62e-1 | 7.38e-2 | 7.30e-2 | 1.66e-2 | **0.00** | **0.00** | **0.00** | **0.00** | **0.00** | **0.00** | 0.00 | 0.00 |
| 4 | **0.00** | **0.00** | 2.85e-1 | 4.57e-2 | 4.32e-2 | 1.94e-2 | **0.00** | **0.00** | **0.00** | **0.00** | **0.00** | **0.00** | 0.00 | 0.00 |
| 5 | **0.00** | **0.00** | 3.61e-1 | 1.45e-1 | 4.69e-2 | 4.22e-2 | 7.64e-3 | 1.91e-2 | **0.00** | **0.00** | **0.00** | **0.00** | 3.31e-5 | 2.27e-4 |
| 6 | 4.41e-4 | 1.64e-3 | 3.25e-1 | 9.50e-2 | 3.19e-2 | 1.10e-2 | 1.06e-2 | 1.65e-2 | **0.00** | **0.00** | **0.00** | **0.00** | 0.00 | 0.00 |
| 7 | **0.00** | **0.00** | 2.88e-1 | 1.04e-1 | 6.53e-2 | 1.41e-2 | 5.61e-3 | 1.49e-2 | **0.00** | **0.00** | **0.00** | **0.00** | 0.00 | 0.00 |
| 8 | **0.00** | **0.00** | 3.07e-1 | 5.18e-2 | 3.46e-2 | 2.12e-2 | **0.00** | **0.00** | **0.00** | **0.00** | **0.00** | **0.00** | 0.00 | 0.00 |

First, we analyze the performance of On-RL. The PPO algorithm possesses strong learning abilities and highly stable convergence properties, which is also the primary reason we select it to generate the expert policy in each environment. The expert policies effectively restrict the frequency deviation of each environment to a very small range, with almost no violations occurring. However, as demonstrated by the performance of On-RL, significant limitations arise when a single PPO policy is required to handle diverse environments. Because it struggles to discern the distinct characteristics of each environment from limited observations, it learns an averaged policy during joint training across the 8 environments. Consequently, the control performance in each individual environment is degraded, and in some cases, its frequency deviation is even an order of magnitude higher than that of the expert policy. On the other hand, it should be emphasized that in real-world microgrids, online learning for RL agents is practically unfeasible due to the excessively high trial-and-error risks. Therefore, training is restricted to offline methods utilizing expert historical trajectories, such as the Off-RL, BC, and DT approaches discussed below.

Second, we evaluate the performance of Off-RL and BC, which are two mainstream offline training mechanisms. They utilize SAC to extract policies from historical data and employ imitation learning to approximate expert policies, respectively. However, similar to On-RL, Off-RL and BC lack the ability to discriminate between environments, making it difficult to achieve efficient control across multiple scenarios. Moreover, due to the inherent instability of Off-RL, it yields the worst performance among all methods, resulting in severe frequency deviations and violations. In real-world microgrids, such fluctuations and violations not only compromise operational stability but can also lead to trip-off of renewable energy sources or even entire system collapse in some extreme cases.

Third, we observe the performance of DT. As can be seen, DT achieves improved performance across all environments compared to Off-RL and BC. This is attributed to two main factors. On the one hand, unlike MDPs that rely on the Markov property, DT formulates the decision-making problem as an autoregressive sequence modeling task. This naturally provides DT with long-range dependencies, enabling it to perceive more complex dynamics and richer information. On the other hand, the Transformer architecture and the larger number of parameters in DT endow it with exceptional learning capabilities, granting it a stronger imitation ability than Off-RL and BC based on traditional neural networks. This paradigm shift from decision-making to sequence generation provides a novel solution pathway for power system dispatch and control.

However, relying solely on the capabilities of DT itself is insufficient. For instance, DT still exhibits constraint violations across several environments. Considering this challenge, some existing methods introduce an adapter head that takes the DT's actions and environmental characteristic parameters as inputs, projecting the original actions from the DT into the specific range of the corresponding environment. Yet, as highlighted previously, many environmental characteristic parameters cannot be explicitly obtained. To circumvent this limitation, we adopt an alternative approach, i.e., extending DT into Prompt-DT and concatenating an expert trajectory specific to the target environment as a prompt before the target sequence generation. As can be seen, this simple yet effective operation yields a substantial performance leap for Prompt-DT over the baseline DT, approaching the experts' decision-making level across many environments. This enables the paradigm of "single-training and multi-site deployment" of the policy network.

Finally, we compare Prompt-DT with the proposed method with context-aware training mechanism. As can be seen, in the vast majority of environments, they achieve highly comparable results and closely approximate the experts' performance. However, in challenging scenarios, such as the second microgrid environment with low inertia, the performance of Prompt-DT is significantly inferior to that of the proposed method. In this environment, the average frequency deviation of Prompt-DT is 275% higher than that of the proposed method, and its frequency violations are roughly 400% greater. This validates the effectiveness and superiority of the proposed context-aware training mechanism, which significantly improves prompt utilization efficiency through feature extraction and self-supervised contrastive learning.

TABLE IV

AVERAGE FREQUENCY DEVIATIONS WITH AND WITHOUT PHYSICS-INFORMED PROMPT DESIGN TECHNIQUE

| Env. ID | Proposed | Proposed w/o Prompt Design | Performance Degradation w/o Prompt Design |
|---|---|---|---|
| 1 | 3.47e-3 | 3.65e-3 | -5.19% |
| 2 | 9.22e-3 | 9.50e-3 | -3.04% |
| 3 | 2.68e-3 | 2.73e-3 | -1.87% |
| 4 | 1.65e-3 | 1.93e-3 | -17.0% |
| 5 | 4.93e-3 | 5.14e-3 | -4.26% |
| 6 | 3.14e-3 | 3.16e-3 | -0.637% |
| 7 | 3.01e-3 | 3.15e-3 | -4.65% |
| 8 | 3.74e-3 | 3.70e-3 | +1.07% |

As for the proposed physics-informed prompt design technique, we also conduct a corresponding ablation study to evaluate the frequency control performance of the proposed method with and without the prompt design technique, and the results are listed in Table IV. It can be observed that when randomly selected prompts are adopted, the performance degrades accordingly. The average frequency deviation increases by approximately 4.5%, and in the fourth environment, the degradation reaches 17%. And in the second challenging environment, the frequency violation increases by nearly 70%. This demonstrates that prompt selection is also a critical component of execution, and a well-crafted prompt can effectively ensure the execution performance of the Prompt-DT.

## *C. Performance on Unseen Environments*

In this subsection, we evaluate the policy's performance on unseen environments and compare three finetuning approaches, i.e., full-parameter, LoRA, and the proposed lightweight method, from three perspectives: the number of trainable parameters, the required number of finetuning samples, and the performance after finetuning.

TABLE V

PARAMETERS OF DIFFERENT FINETUNING APPROACHES

| Finetuning Approach | Trainable Parameters | Proportion |
|---|---|---|
| Full-Parameter | 1,332,356 | 100% |
| LoRA (Rank=8) | 230,916 | 17.3% |
| Proposed | 132,612 | 9.95% |

First, the number of parameters involved in each finetuning approach is listed in Table V. Specifically, full-parameter finetuning requires adjusting all parameters, which imposes a substantial computational burden and communication overhead on the microgrid operator. The mainstream LoRA finetuning effectively reduces the number of trainable parameters to approximately 17% by freezing the main model parameters and inserting trainable low-rank matrices. In contrast, our proposed method freezes the LLM backbone and only adjusts the parameters of selected heads, further reducing the proportion of trainable parameters to about 10%. This not only alleviates computational cost but also offers greater flexibility, as it avoids contaminating the LLM backbone.

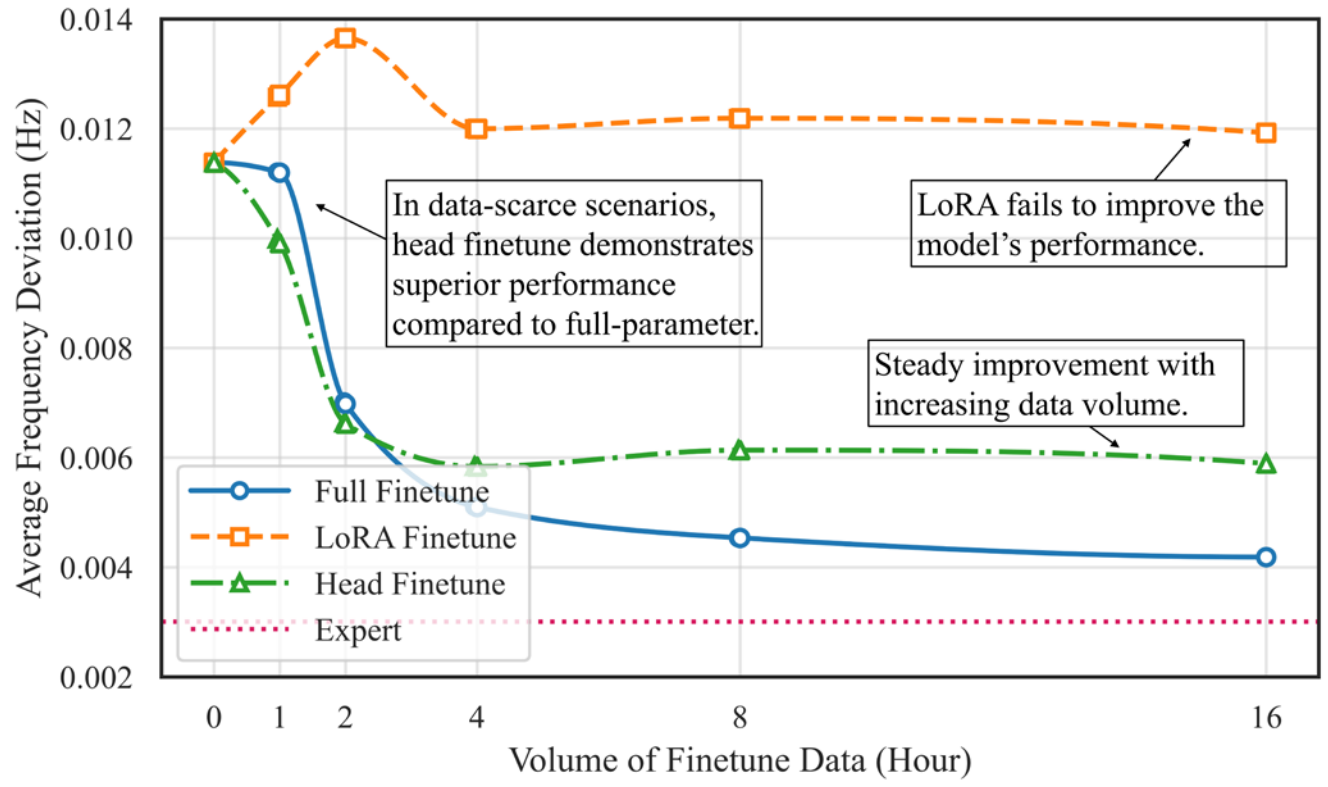

**Fig. 5.** Frequency control performance using different finetuning approaches in the unseen environment.

Subsequently, we collect several hours of expert trajectories on the target environment as finetuning data, and finetune the pretrained model using trajectories of 1, 2, 4, 8, and 16 hours to observe the performance of each approach, which is presented in Fig. 5. As can be seen, without finetuning, the model yields unsatisfactory performance, reaching an average frequency deviation of over 0.01 Hz, which also corroborates the necessity of the finetuning procedure. With only two hours of expert trajectories, the model performance can be effectively improved, and it steadily improves and converges as the number of samples increases, demonstrating the generalization capability of the Prompt-DT. From the comparison of the three methods, it can be observed that LoRA performs the worst, with almost no improvement or even a slight degradation after finetuning. When the number of finetuning samples is small, the proposed approach even slightly outperforms full-parameter finetuning, because it preserves the general knowledge learned by the LLM backbone to the greatest extent. When the number of finetuning samples increases, full-parameter finetuning achieves the best performance, which is a reasonable outcome since it can explore a larger space of parameter combinations.

At last, one of the primary concerns introduced by the LLM backbone is the inference latency. Therefore, we measure the one-step inference latency of each method and summarize the results in Table VI for readers' reference.

TABLE VI

ONE-STEP INFERENCE LATENCY OF DIFFERENT METHODS

| Method | Inference Latency |
|---|---|
| On-RL | 0.190 ms |
| Off-RL | 0.381 ms |
| BC | 0.244 ms |
| DT | 4.29 ms |
| Prompt-DT | 4.69 ms |
| Proposed | 4.77 ms |
| Expert | 0.174 ms |

As can be seen from the results listed in Table VI, due to the adoption of the Transformer architecture, DT, Prompt-DT, and the proposed method exhibit inference latencies that are an order of magnitude higher than traditional neural networks. Meanwhile, because Prompt-DT and the proposed method prepend a prompt before the target sequence, their latencies are slightly higher than that of DT, which is essentially attributed to the increased sequence length. However, all methods can complete one-step inference within 5 ms, thus having virtually no impact on second-level and minute-level control problems.

## V. Conclusion

To address the generalization challenge of existing models, we innovatively introduce Prompt-DT and propose context-aware training and execution mechanism, along with a lightweight finetuning approach. Our proposed method successfully solves the microgrid frequency control problem across diverse scenarios and configurations, enabling the paradigm of "single-training and multi-site deployment", and can be transferred to unseen environments with minimal samples and minimal computational requirements.

As this constitutes a fundamentally novel dispatch and control paradigm, in future work, its safety, robustness, and scalability warrant further and more in-depth investigations.

## References

[1] Y. Zhang *et al.*, "A Temperature-Aware Operating Model of EV Battery for Resilient Residential Microgrids in Frigid Weather," *IEEE Trans. Smart Grid*, vol. 17, no. 1, pp. 483-496, Jan. 2026.

[2] M. Macmillan, A. Zolan, M. Bazilian, and D. L. Villa, "Microgrid design and multi-year dispatch optimization under climate-informed load and renewable resource uncertainty," *Applied Energy*, vol. 368, no. 123355, 2024.

[3] Z. Li, Z. Cheng, J. Si, and S. Xu, "Distributed Event-Triggered Hierarchical Control of PV Inverters to Provide Multi-Time Scale Frequency Response for AC Microgrid," *IEEE Trans. Power Systems*, vol. 38, no. 2, pp. 1529-1542, Mar. 2023.

[4] W. Tang, X. Nie, T. Han, T. Qian, and Y. Xue, "Distributed dynamic event-triggered secondary control in dually heterogeneous islanded PV–BESS microgrids for voltage–frequency regulation and SoC balancing," *IEEE Trans. Smart Grid*, early access.

[5] K. Pokharel, H. Li, and H. He, "World-Model-Based Deep Reinforcement Learning for Real-Time Microgrid Energy Management," *IEEE Trans. Smart Grid*, early access.

[6] J. Yang, W. Huang, Y. Feng, Z. Jin, Z. Shuai, and Z. J. Shen, "Frequency Response Model and Inertia Enhancement Method of Islanded Microgrid Integrated With Free Piston Stirling Generator," *IEEE Trans. Sustainable Energy*, vol. 17, no. 3, pp. 3006-3018, Jul. 2026.

[7] W. Breesam, R. Alamian, N. Tashakor, B. E. Youcefa, and S. M. Goetz, "Frequency Control in Microgrids: A Fuzzy Neural Network-Based Adaptive Virtual Synchronous Generator," *IEEE Trans. Smart Grid*, vol. 17, no. 3, pp. 1782-1793, May 2026.

[8] X. Yang, H. Liu, and W. Wu, "Attention-Enhanced Multi-Agent Reinforcement Learning Against Observation Perturbations for Distributed Volt-VAR Control," *IEEE Trans. Smart Grid*, vol. 15, no. 6, pp. 5761-5772, Nov. 2024.

[9] Y. Ma, Z. Hu, and Y. Song, "A Reinforcement Learning Based Coordinated but Differentiated Load Frequency Control Method With Heterogeneous Frequency Regulation Resources," *IEEE Trans. Power Systems*, vol. 39, no. 1, pp. 2239-2250, Jan. 2024.

[10] X. Chen, M. Zhang, Z. Wu, L. Yu, N. D. Hatziargyriou, and X. Guan, "Load Frequency Control of Multi-Microgrids Based on Deep Deterministic Policy Gradient Integrated With Online Learning," *IEEE Trans. Smart Grid*, vol. 16, no. 5, pp. 4266-4278, Sep. 2025.

[11] R. Wang, S. Bu, and C. Y. Chung, "Real-Time Joint Regulations of Frequency and Voltage for TSO-DSO Coordination: A Deep Reinforcement Learning-Based Approach," *IEEE Trans. Smart Grid*, vol. 15, no. 2, pp. 2294-2308, Mar. 2024.

[12] X. Wan and M. Sun, "AdapSafe2: Prior-Free Safe-Certified Reinforcement Learning for Multi-Area Frequency Control," *IEEE Trans. Power Systems*, vol. 40, no. 3, pp. 2244-2257, May 2025.

[13] A. S. Mohamed and D. Kundur, "On the Use of Reinforcement Learning for Attacking and Defending Load Frequency Control," *IEEE Trans. Smart Grid*, vol. 15, no. 3, pp. 3262-3277, May 2024.

[14] L. Zeng and M. Sun, "Bridge the Sim-to-Real Gap in Virtual Synchronous Generator-Based Frequency Control With Robust Deep Reinforcement Learning," *IEEE Trans. Power Systems*, vol. 41, no. 4, pp. 2682-2694, Jul. 2026.

[15] Y. Pan, J. Hu, Y. Fang, and N. Liu, "Robust Optimization and DRL-Based Control Strategies for Electric Vehicle Aggregators to Provide Frequency Regulation Service," *IEEE Trans. Smart Grid*, vol. 16, no. 6, pp. 5262-5274, Nov. 2025.

[16] M. Zhang, Y. Xu, L. Sang, Y. Guo, and N. D. Hatziargyriou, "Virtual Power Plants for Frequency Regulation: A Learning-Based Method With Safety Guarantee," *IEEE Trans. Smart Grid*, vol. 17, no. 1, pp. 388-401, Jan. 2026.

[17] T. Zhao, A. Yogarathnam, and M. Yue, "A Large Language Model for Determining Partial Tripping of Distributed Energy Resources," *IEEE Trans. Smart Grid*, vol. 16, no. 1, pp. 437-440, Jan. 2025.

[18] X. Yang *et al.*, "Large Language Model Powered Automated Modeling and Optimization of Active Distribution Network Dispatch Problems," *IEEE Trans. Smart Grid*, vol. 17, no. 2, pp. 952-965, Mar. 2026.

[19] A. Jena, F. Ding, J. Wang, Y. Yao, and L. Xie, "LLM-Based Adaptive Distribution Voltage Regulation Under Frequent Topology Changes: An In-Context MPC Framework," *IEEE Trans. Smart Grid*, vol. 16, no. 5, pp. 4297-4300, Sep. 2025.

[20] X. Yang, C. Lin, H. Liu, and W. Wu, "RL2: Reinforce Large Language Model to Assist Safe Reinforcement Learning for Energy Management of Active Distribution Networks," *IEEE Trans. Smart Grid*, vol. 16, no. 4, pp. 3419-3431, Jul. 2025.

[21] X. Yang *et al.*, "Two-Stage Active Distribution Network Voltage Control via LLM-RL Collaboration: A Hybrid Knowledge-Data-Driven Approach," *IEEE Trans. Smart Grid*, early access.

[22] Z. Li *et al.*, "OptDisPro: LLM-Based Multi-Agent Framework for Flexibly Adapting Heuristic Optimal DisFlow," *IEEE Trans. Smart Grid*, vol. 17, no. 1, pp. 794-805, Jan. 2026.

[23] M. Lin, T. Wu, and K. Xie, "Low-Carbon Scheduling for Power-Hydrogen Integrated Energy System Using Large Language Model Enhanced Deep Reinforcement Learning," *IEEE Trans. Sustainable Energy*, early access.

[24] X. Liu, B. Sun, and H. Yu *et al.*, "Two-stage fine-tuning of large language models for distribution network service restoration via CoT reasoning," *Applied Energy*, vol. 423, no. 128417, 2026.

[25] C. Liu, Y. Wang, Y. Zou, and Q. Cui, "Text-to-control of networked microgrids: A large language model agent workflow for reinforcement learning scheduling," *Applied Energy*, vol. 422, no. 128289, 2026.

[26] M. Zhang, M. Liu, H. Wang, Y. Wen, A. -L. Luo, and Y. Zhang, "Leveraging Large Language Model for Generalization in Building Energy Management," *IEEE Trans. Smart Grid*, vol. 16, no. 6, pp. 4712-4725, Nov. 2025.

[27] L. Chen, K. Lu, and A. Rajeswaran *et al.*, "Decision transformer: Reinforcement learning via sequence modeling," *Advances in Neural Information Processing Systems*, no. 34, pp. 15084-15097, 2021.

[28] Y. Xia *et al.*, "A Safe Policy Learning-Based Method for Decentralized and Economic Frequency Control in Isolated Networked-Microgrid Systems," *IEEE Trans. Sustainable Energy*, vol. 13, no. 4, pp. 1982-1993, Oct. 2022.

[29] M. Xu, Y. Shen, and S. Zhang *et al.*, "Prompting decision transformer for few-shot policy generalization," in *International Conference on Machine Learning*, pp. 24631-24645, 2022.

[30] E. J. Hu, Y. Shen, and P. Wallis *et al.*, "Lora: Low-rank adaptation of large language models," *arXiv:2106.09685*, Oct. 2021.

[31] R. S. Sutton and A. G. Barto, *Reinforcement Learning: An Introduction*, vol. 1. Cambridge, MA, USA: MIT Press, 1998.

[32] A. Vaswani, N. Shazeer, and N. Parmar *et al.*, "Attention is all you need," *Advances in Neural Information Processing Systems*, no. 30, 2017.

[33] X. Yang *et al.*, "Supplementary files for LLMs are Few-Shot Decision-Makers: Generalized Context-Aware Microgrid Frequency Control through Prompt Decision Transformer," Aug. 2026, [Online]. Available: https://github.com/YangXuSteve/Prompt-Decision-Transformer.

[34] J. Schulman, F. Wolski, P. Dhariwal, A. Radford, and O. Klimov, "Proximal policy optimization algorithms," *arXiv:1707.06347*, Aug. 2017.

[35] T. Haarnoja, A. Zhou, and K. Hartikainen *et al.*, "Soft actor-critic algorithms and applications," *arXiv:1812.05905*, Jan. 2019.